\documentclass[11pt, aps]{revtex4}
\usepackage{graphicx}
\usepackage{amssymb}
\usepackage{amsmath}
\usepackage{latexsym}

\begin{document}

\title{A hemispherical, high-solid-angle optical micro-cavity for cavity-QED studies}

\author{Guoqiang Cui, J. M. Hannigan, R. Loeckenhoff, F. M. Matinaga, M. G. Raymer}
\affiliation{Oregon Center for Optics and Department of Physics\\
University of Oregon, Eugene, OR 97403 USA }
\email{raymer@uoregon.edu}

\author{S. Bhongale, M. Holland}
\affiliation{JILA, University of Colorado, Boulder, CO 80309 USA}
\email{mholland@bdagger.colorado.edu}

\author{S. Mosor, S. Chatterjee, H. M. Gibbs, G. Khitrova}
\affiliation{Optical Science Center, University of Arizona, Tucson,
AZ 85721 USA} \email{hyattgibbs@att.net}

\begin{abstract}
We report a novel hemispherical micro-cavity that is comprised of a
planar integrated semiconductor distributed Bragg reflector (DBR)
mirror, and an external, concave micro-mirror having a radius of
curvature $50\,\mathrm{\mu m}$. The integrated DBR mirror containing
quantum dots (QD), is designed to locate the QDs at an antinode of
the field in order to maximize the interaction between the QD and
cavity. The concave micro-mirror, with high-reflectivity over a
large solid-angle, creates a diffraction-limited (sub-micron)
mode-waist at the planar mirror, leading to a large coupling
constant between the cavity mode and QD. The half-monolithic design
gives more spatial and spectral tuning abilities, relatively to
fully monolithic structures. This unique micro-cavity design will
potentially enable us to both reach the cavity quantum
electrodynamics (QED) strong coupling regime and realize the
deterministic generation of single photons on demand.
\end{abstract}

\maketitle

\section{Introduction}

Optical micro-cavities have played a central role in achieving
strong coupling between a single atom and a mode of an optical
cavity, which enables a range of novel phenomena that rely on the
control of the mode structure of the vacuum (so-called cavity-QED
effects). These include enhanced or suppressed spontaneous emission
\cite{Purcell46, Hulet85, DeMartini87, Heinzen871, Heinzen872,
Morin94}, thresholdless lasing \cite{Yamamoto93, Rice94},
normal-mode splitting \cite{Kimble94}, and optical nonlinearity at
the single-photon level \cite{Turchette95}. In the last two decades,
such strong coupling has been achieved in free-space atomic systems,
such as a dilute atomic beam passing through a short
($10-100\,\mathrm{\mu m}$ length) optical cavity \cite{Meschede85,
Brune87, Rempe91}, or through a cold microwave cavity
\cite{Haroche89}.

There is also interest in achieving strong cavity-QED coupling in
semiconductor QD systems, following early studies using planar
quantum-well-cavity systems, which themselves cannot reach this
regime \cite{Khitrova99}. Recent experiments showed signatures of
strong coupling in some monolithic structures such as micro-pillar
\cite{Reithmaier04}, photonic crystal nanocavity \cite{Yoshie04} and
micro-disk \cite{Peter05}. Obvious advantages of using QDs in such
schemes are that the QDs are stationary and they exist in a
solid-state system, which can be optically or electrically pumped
\cite{Yuan02}. The principal disadvantages in these monolithic
structures, however, are the lack of efficient control of the
spatial and spectral overlap between QDs resonance and cavity modes.
For instance, temperature tuning of the QD has to be used to tune
through cavity resonance \cite{Reithmaier04, Yoshie04, Peter05},
which is undesirable because the dipole dephasing rate increases at
elevated temperatures {\cite{Fan98, Bayer02}.

This report focuses on the design, modeling, fabrication, and
performance of a unique half-monolithic, hemispherical micro-cavity
for semiconductor cavity-QED. The cavity parameters are in a novel
range: cavity length $= 40-60\,\mathrm{\mu m}$, finesse $= 200$
(which should be amenable to increase by an order of magnitude),
mode-waist size $\approx 1\,\mathrm{\mu m}$, mode divergence angle
$\pm40$ deg. This cavity design contains two unique features---the
use of a concave micro-mirror with high-reflectivity over a
large-solid angle and the use of an integrated DBR mirror containing
the QD sample in an external-cavity configuration. The 40--60-micron
curved mirror substrate has a high degree of sphericity and an
excellent surface quality, enabling the application of a
custom-designed multilayer dielectric coating with 99.5\%
reflectivity over a high-solid-angle \cite{Coating}. Such large
solid angle is unique compared with, for example, a recently
reported half-monolithic micro-cavity design \cite{Trupke05}.

One potential application of such a cavity/QD system is for
semiconductor cavity-QED study; the other is the generation, on
demand, of single photons or of photon pairs. The cavity can also be
operated with a standard planar dielectric mirror replacing the
semiconductor DBR mirror. Such an all-dielectric cavity may find
uses in atomic cavity-QED or cold-atom studies, or in novel forms of
microscopy or interferometry.

The cavity components have been fabricated in our collective
laboratories---the concave micro-mirror by a novel gas-bubble
technique and the DBR/QD structure by molecular-beam epitaxy (MBE).

\section{Cavity Design Overview}

Figure \ref{Fig. 1} shows a real structure and a schematic diagram
of the cavity. A transparent, planar substrate with a multilayer DBR
coating (made either of semiconductors or optical coating
dielectrics) forms one end of the cavity. A transparent concave
glass surface with a dielectric multilayer reflective coating forms
the other end. In between is air or vacuum. The radius of curvature
of the mirror is denoted $\mathrm{R_M}$, and can be fabricated in
the range $40-100\,\mathrm{\mu m}$. The on-axis distance L between
the surfaces of the two mirrors is referred to as the cavity length.
In a hemispherical cavity these lengths are equal, $\mathrm{L=R_M}$.
This places the cavity on the boundary for stability, and (in the
paraxial approximation, which actually fails here) leads to the
interesting property that the modes fall into groups with a high
degree of frequency degeneracy \cite{Siegman86}.

\begin{figure}[htbp]
  \centering
  \begin{minipage}[b]{0.23\textwidth}
    \centering
    \includegraphics[width=0.95\textwidth]{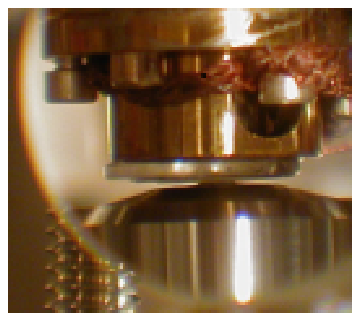}
  \end{minipage}
  \begin{minipage}[b]{0.63\textwidth}
    \centering
    \includegraphics[width=0.95\textwidth]{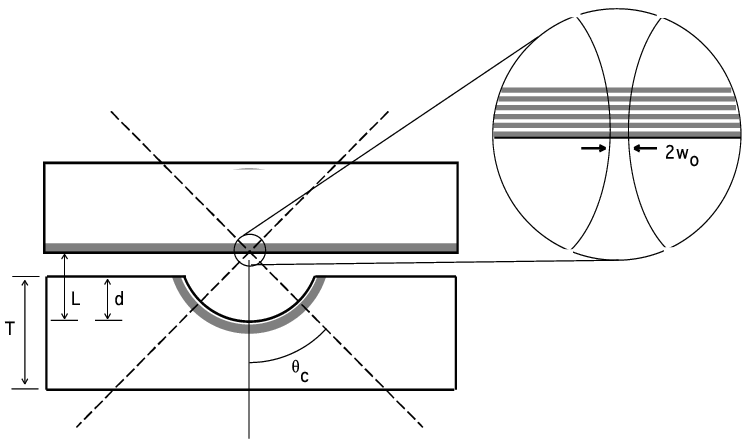}
  \end{minipage}
\caption{Hemispherical cavity, comprised of a planar substrate and a
concave glass surface with layer reflective coating (shown as grey
region). The dashed lines approximate the 1/e intensity contours of
the fundamental mode in the cavity and its continuation outside. The
angular half-width of the mode is $\mathrm{\theta_C}$. The blow-up
shows the DBR and the mode contours in the waist region. Typically
the length L is $50\,\mathrm{\mu m}$, the depth d is
$30\,\mathrm{\mu m}$ and the waist diameter is $2w_0=1\,\mathrm{\mu
m}$.} \label{Fig. 1}
\end{figure}

The radius of the mode waist, located at the planar mirror, is
denoted $w_0$. Since the QD is to be placed in this waist, this
radius should be minimized in order to maximize the coupling between
the QD and the field. The angular half-width of the cavity mode is
$\mathrm{\theta_C}$. Diffraction dictates that the smaller $w_0$ is
made, the larger $\mathrm{\theta_C}$ becomes. When $w_0$ equals one
optical wavelength, the angle $\mathrm{\theta_C}$ is roughly 40 deg.
For such large angles, the electromagnetic field cannot be
completely transverse to the cavity axis, as would be the case in
the paraxial limit where $\mathrm{\theta_C}$ is restricted to very
small values. This indicates a need for a theory, summarized below,
beyond the common paraxial treatment.

The effective mode volume $\mathrm{V_{eff}}$, which depends on the
location of the QD, is defined as spatial integral of the field
intensity, normalized to unity at the maximum. For example, if the
mode amplitude can be described as a (paraxial) Gaussian function
with 1/e amplitude contours that define a spot size $w(z)$ at the
position \textit{z} along the cavity axis, then the effective mode
volume is given by $\mathrm{V_{eff}}=\pi w_0^2\mathrm{L}/4$
\cite{Drummond81}. It is thus important to locate the QD at an
antinode of the field at the waist.

\subsection{Concave Micro-mirror Substrate}

A unique component of our cavity is the concave micro-mirror. We
developed a technique for its in-house fabrication. For use in a
high-finesse cavity, it is crucial that the curved surface of the
mirror substrate be smooth on nanometer scales. This prevents undue
amounts of light scattering that would act as a loss, spoiling the
finesse.

Our technique, shown in Fig. \ref{Fig. 2}(a), proceeds by melting a
stack of small, high-quality borosilicate glass tubes under a
nitrogen atmosphere, trapping small gas bubbles. By surface tension
the gas bubbles are naturally created with high degree of
sphericity. After the glass cools and hardens, we grind and polish
it on a simple optical polishing wheel so that about one-third of a
selected bubble remains embedded in the surface. The top surface,
where a few bubbles are open, is finished with diamond disc
featuring nickel-plated diamonds in a raised dot matrix pattern of
$6\,\mathrm{\mu m}$ grit size on a polishing wheel. The bottom
surface is polished using a $0.05\,\mathrm{\mu m}$ colloidal silica
suspension on a polishing cloth, to achieve an optical-quality
finish. Finally, we obtain a flat sample of about
$\mathrm{T=150\,\mu m}$ thickness, which forms our concave mirror
substrate.

\begin{figure}[htbp]
  \centering
  \begin{minipage}[b]{0.7\textwidth}
    \centering
    \includegraphics[width=0.9\textwidth]{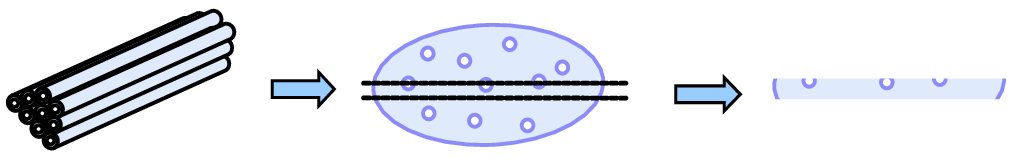}\\
    {(a)}
  \end{minipage}
  \begin{minipage}[b]{0.18\textwidth}
    \centering
    \includegraphics[width=0.9\textwidth]{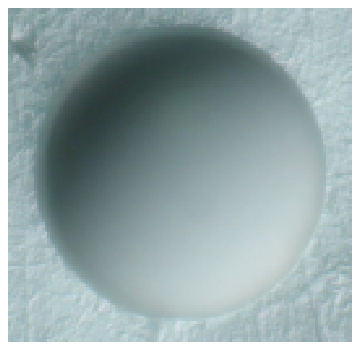}\\
    {(b)}
  \end{minipage}
\caption{(a) Melting borosilicate glass tubes to form nitrogen
gas-bubbles in the glass and polishing the glass bulk into a
$150\,\mathrm{\mu m}$-thick slide. (b) 40X pictures of a dimple.
Diameter of the dimple $=200\,\mathrm{\mu m}$.}
  \label{Fig. 2}
  \end{figure}

Figure \ref{Fig. 2}(b) shows images of a typical dimple at 40X
magnification. The planar surface on the top side, surrounding the
dimples, is very rough, as a result of the final $6\,\mathrm{\mu
m}$-grit used on this side. This was chosen to minimize the amount
of contaminating sub-micron glass dust produced during polishing.
The inside of the dimple (out of focus here) is far smoother. The
dimples will ideally have an opening half-angle of
$\mathrm{\theta_C}\approx 40$ deg, a radius of curvature of
$\mathrm{R_M}\approx 50\,\mathrm{\mu m}$ and a surface with sub-nm
roughness.

We expected a good sphericity of the dimple surfaces since for
decreasing dimensions the surface tension is an increasingly strong
force compared to other forces like gravity. The sphericity has been
measured with a Wyko interferometer, see Fig. \ref{Fig. 3}. At the
bottom of a dimple, in a circle of $15\,\mathrm{\mu m}$ diameter,
the deviations from perfect sphericity where found to be less than
$10\,\mathrm{nm}$.

\begin{figure}[htbp]
\centering
\includegraphics[width=0.85\textwidth]{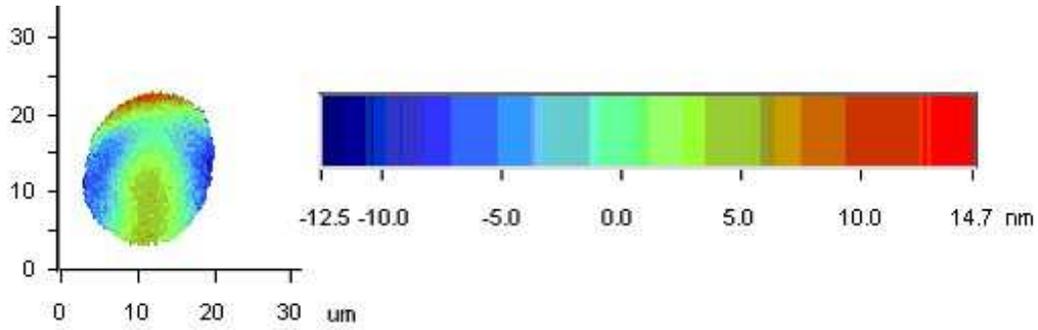}
\caption{Measured sphericity of the dimple with a Wyko
interferometer at the University of Arizona.} \label{Fig. 3}
\end{figure}

The surface roughness was also measured using a Wyko interferometer
that carries out a Fourier-analysis of the surface to determine the
power (spatial) spectral density (PSD) of surface roughness as a
function of the lateral size of the errors. Figure \ref{Fig. 4}(a)
shows the measured PSD of five dimples.

\begin{figure}[htbp]
\centering
\begin{minipage}[b]{0.44\textwidth}
\centering
\includegraphics[width=0.95\textwidth]{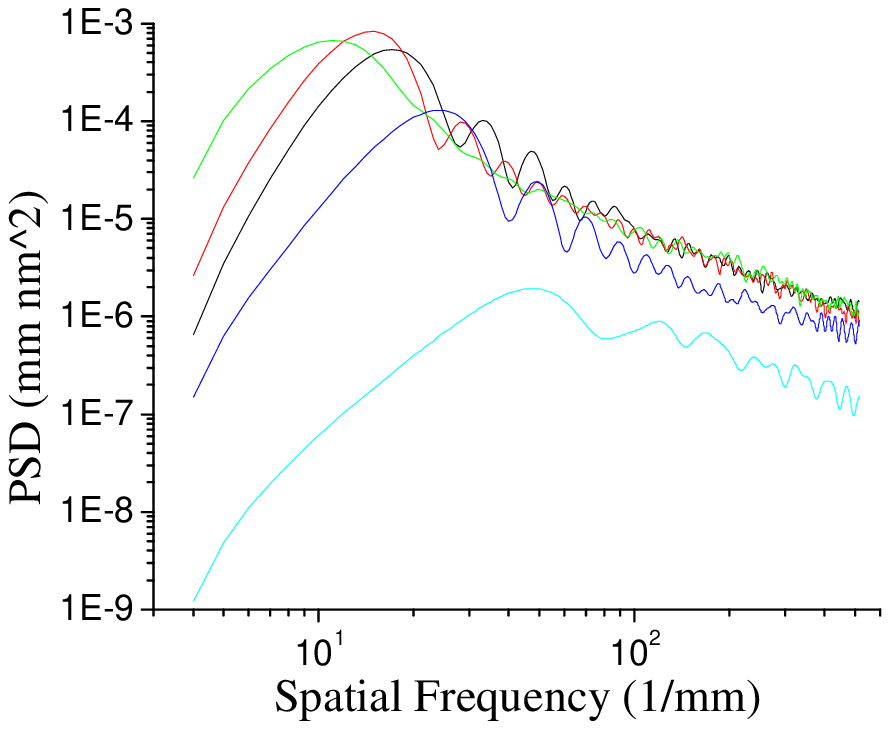}\\
{(a)}
\end{minipage}
\begin{minipage}[b]{0.45\textwidth}
\centering
\includegraphics[width=0.95\textwidth]{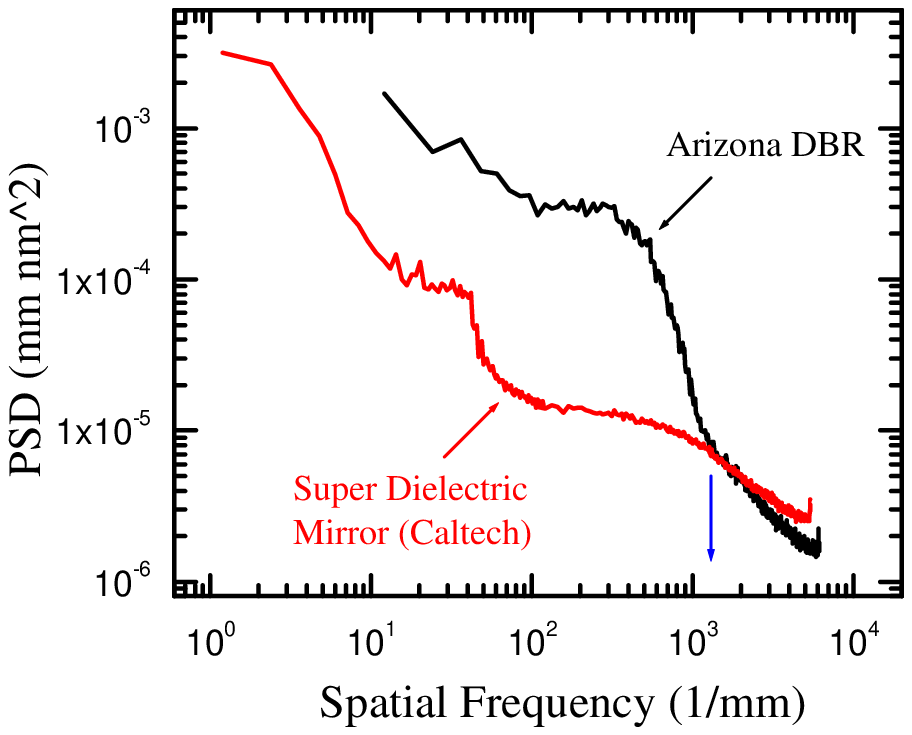}\\
{(b)}
\end{minipage}
\caption{(a) Measured PSD surface roughness for five dimples and (b)
semiconductor DBR mirror and super dielectric mirror with a Wyko
interferometer. The relevant length scale (indicated by the blue
arrow) is about one micron because our unique cavity design yields a
waist size at the DBR of this size.} \label{Fig. 4}
\end{figure}

For errors with a transverse spatial frequency greater than
$300\,\mathrm{mm^{-1}}$, the surface quality competes with the best
planar mirrors available, see Fig. \ref{Fig. 4}(b). However the
roughness increases dramatically for smaller spatial frequencies
(large length scales). We are not sure whether this represents
intrinsic errors like wrinkles formed in the cooling process or
debris left from polishing.

\subsection{Optical Coating for Curved Micro-Mirror}

Optical coating of such a small and highly curved dimple substrate
is a nonstandard procedure. One problem is that the atomic coating
beam is incident on the curved substrate at a different angle at
each different location. This alters the deposition rate in a
location-dependent manner, which leads to systematic variation of
the layer thickness and therefore of edge wavelengths of the
coating's stop-band. Therefore, we designed the coating scheme
(using TFCalc), in a way that compensates for the large change of
coating-beam angle across the surface of the substrate.

We designed a high-index-contrast TiO2-SiO2 coating, having a
stop-band shifted to longer wavelength at the center of the dimple.
For a working wavelength of 750 nm, the reflectivity is greater than
95\% between 737 nm and 808 nm. For locations away from the center,
the coating becomes thinner, shifting the stop band to shorter
wavelengths. At some location on the dimple surface (or angle from
the optical axis at the mode focus region), the stop band suddenly
shifts past the working wavelength, causing a sudden drop of mirror
reflectivity, as has also been observed in \cite{Trupke05}.

Measurements, shown in Fig. \ref{Fig. 5}(a), of the dimple-mirror
transmission versus angle from the optical axis confirms that our
design and fabrication has succeeded in giving a high reflectivity
(99.5\% or higher) over a wide angular range of $\pm40$ deg, which
is wide enough to support the hemispherical modes of interest. The
coated curved dimple was then glued, using index-matching optical
adhesive, to the face of a 100X immersion-microscope objective
(Zeiss Plan-NEOFLUAR) with a numerical aperture $\mathrm{NA=1.3}$ in
order for efficient mode-coupling over a high-solid angle, see Fig.
\ref{Fig. 5}(b). To ensure proper positioning of the dimple, we glue
it while monitoring interferometrically by a Twyman-Green setup
\cite{Born99}, in which a laser beam passes into the objective,
reflects from the dimple and interferences with a reference beam.

\begin{figure}[htbp]
\centering
\begin{minipage}[b]{0.45\textwidth}
\centering
\includegraphics[width=0.9\textwidth]{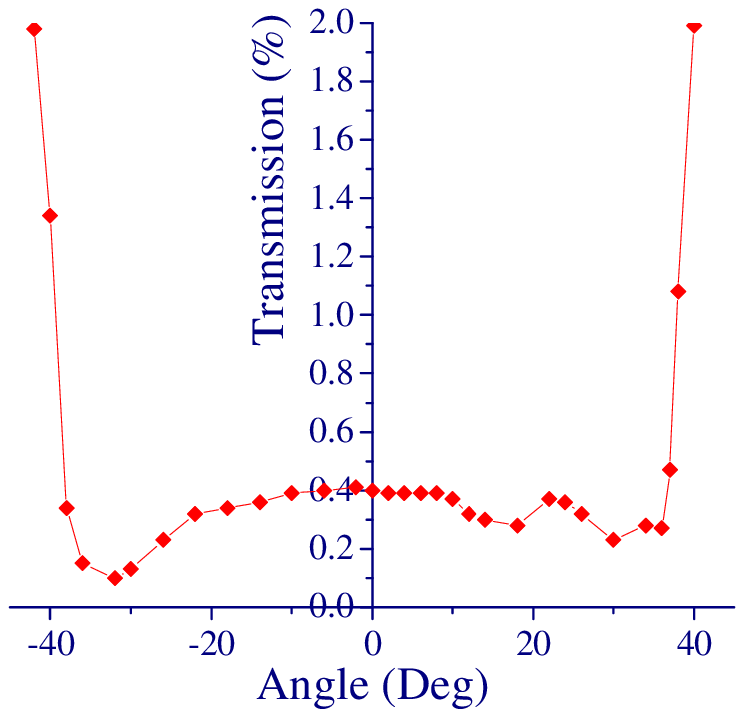}\\
{(a)}
\end{minipage}
\begin{minipage}[b]{0.42\textwidth}
\centering
\includegraphics[width=0.9\textwidth]{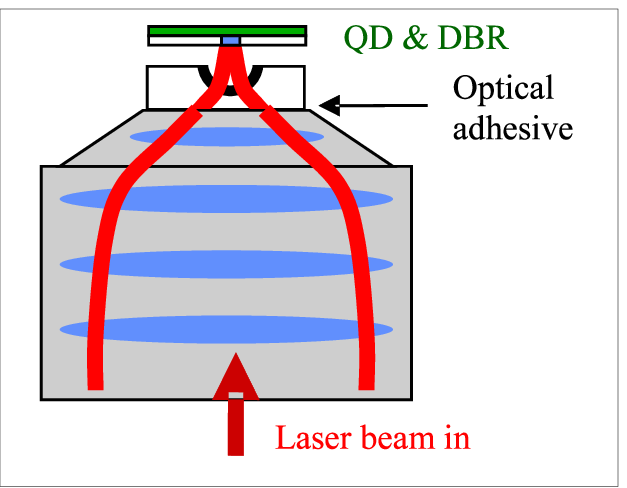}\\[9mm]
{(b)}
\end{minipage}
\caption{(a) Measured dimple-mirror transmission versus angle from
the optical axis at the mode focus region. (b) The coated curved
dimple is glued using index-matching optical adhesive to the face of
a 100X immersion-microscope objective with NA=1.3.} \label{Fig. 5}
\end{figure}

\subsection{MBE-Growth and Characterization of Integrated Top Mirror and QD layer}

Semiconductor planar DBR mirror with exceptionally good surface
smoothness and high reflectivity can be grown by MBE technique
\cite{Stanley94}, We found that the surface roughness on transverse
length scales relevant for our needs ($\sim 1\,\mathrm{micron}$) is
equal to that of the best super-polished dielectric mirrors of the
type used in atomic cavity-QED experiments.

Figure \ref{Fig. 4}(b) shows a comparison of two kinds of
mirrors---the MBE-grown and a commercial super polished dielectric
mirror (made by Kimble group at the Caltech). The figure plots the
PSD of surface roughness versus transverse spatial frequency,
measured with a Wyko interferometer. It is seen that the planar
semiconductor mirror has far larger roughness for low spatial
frequencies, while the commercial dielectric mirror is slightly
rougher at spatial frequencies above $100\,\mathrm{mm^{-1}}$, the
region of interest for our cavity, since the mode waist is less than
$1\,\mathrm{\mu m}$.

\begin{figure}[htbp]
\centering
\includegraphics[width=0.8\textwidth]{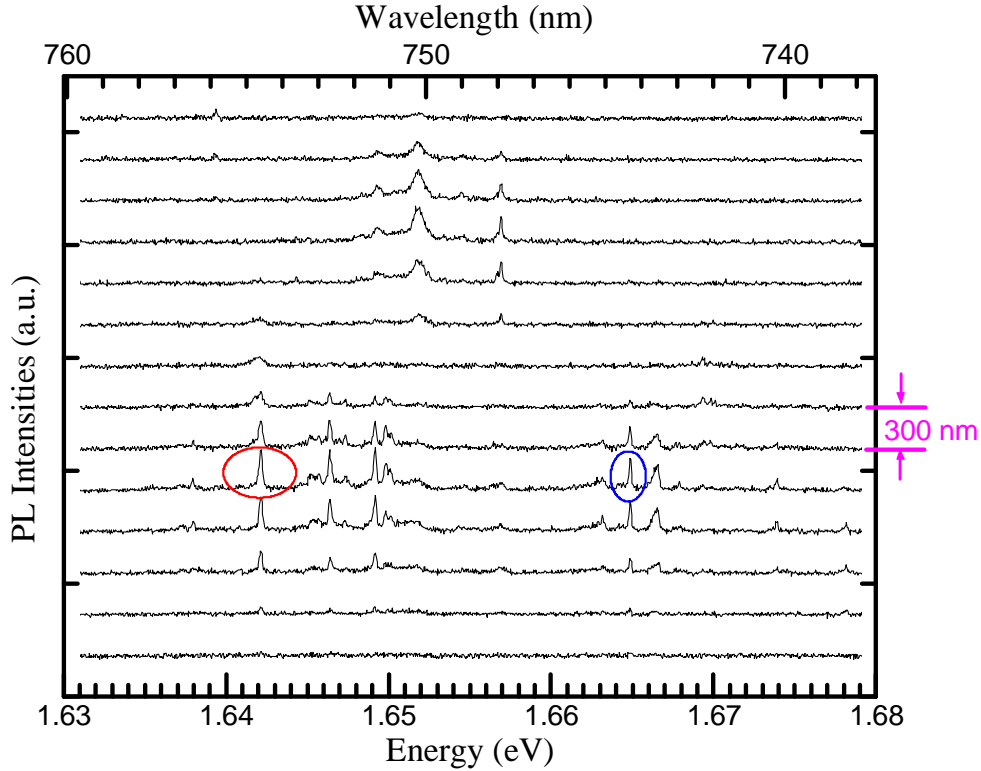}
\caption{Nano-scope spectral scans of different spatial locations on
UA-grown sample, showing both spectrally and spatially well isolated
single QD emission line (red circled) at low temperature in the
750--760 nm target region.} \label{Fig. 6}
\end{figure}

The GaAs QDs that we use are interface-fluctuation quantum dots
(IFQDs). They are formed through the influence of monolayer-thick
interface fluctuation during the MBE-growth of a quantum well (QW),
creating elliptically shaped regions about 50--100 nm across
\cite{Zrenner94, Gammon96}. We have succeeded in growing
good-quality IFQDs on the top surface of high-quality DBRs. The QDs
are embedded in a wavelength-thick spacer layer to place the QDs at
an antinode of the cavity and have a relatively large dipole moment
(60 Debye), enabling them to interact strongly with the cavity
field.

Figure \ref{Fig. 6} shows a set of nano-scope photoluminescence (PL)
spectra \cite{Gammon96} for a sequence of different locations with a
spatial step of 300 nm in the QW plane on such a DBR. The broader PL
emission lines in the upper traces are inhomogeneously broadened and
can be fit approximately by Gaussian distributions, and are likely
caused by emission from spatially close QDs with different lateral
sizes and hence different emission frequencies. Some of the narrower
PL emission lines in the lower traces can be fit by Lorentzian
distributions, which are homogeneously broadened, signifying
isolated single quantum emitters, and can be identified as emissions
from single QDs, for example, the red circled and blue circled ones.
The red circled one is in our 750--760 nm target region and is
spectrally well isolated. The emission lines at the same energy
(wavelength) in the adjacent traces are also from this specific QD
and indicate that it is also spatially well isolated ($\sim
600\,\mathrm{nm}$) from other QDs.

\section{Cavity Construction, Testing and Modeling}

We constructed and tested a high-quality hemispherical cavity using
our 60-micron mirror and a planar semiconductor DBR (CAT 96)
containing QDs located at the center of a one-wavelength spacer
layer. The semiconductor DBR mirror is mounted on a tripod system,
supported by three Burleigh UHVL Inchworm Motors, to control
precisely its longitudinal position and its angle with respect to
the curved mirror. The tripod also contains an \textit{x-y}
nano-positioner, which can laterally scan the mode waist in a
$50\times50\,\mathrm{\mu m^2}$ region, essential for scanning and
addressing a single QD, and a piezoelectric stack driven by a
laser-referenced feedback loop for stabilizing the length of the
cavity. The system operates inside a high-vacuum chamber ($10^{-8}$
torr), to allow cooling the DBR mirror to around 10--17 K to reduce
QD dephasing rates and to avoid coating of the DBR mirror by
cryopumping and attendant absorption and scattering.

\subsection{Cavity-mode structure and finesse}

We tested the cavity by passing laser light through it and observing
the cavity modes and measuring its finesse. Figure \ref{Fig. 7}
shows several well-defined cavity modes observed for different
frequencies. We label them using HG and LG notation, since they are
qualitatively similar to the Hermite-Gauss or Laguerre-Gauss modes
that are applicable in the paraxial limit \cite{Siegman86}.

\begin{figure}[htbp]
\centering
\begin{minipage}[c]{0.2\linewidth}
\centering
\includegraphics[width=0.95\textwidth]{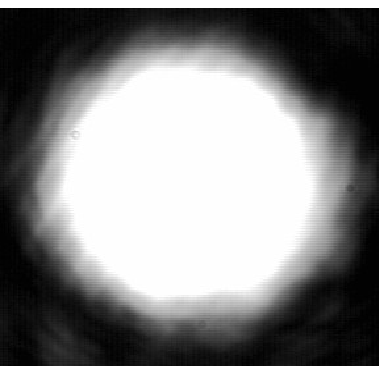}
\end{minipage}
\begin{minipage}[c]{0.2\linewidth}
\centering
\includegraphics[width=0.95\textwidth]{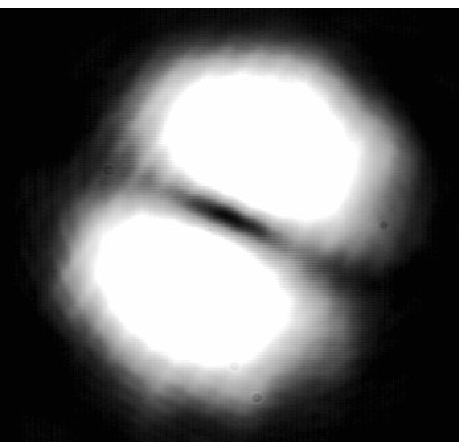}
\end{minipage}
\begin{minipage}[c]{0.2\linewidth}
\centering
\includegraphics[width=0.95\textwidth]{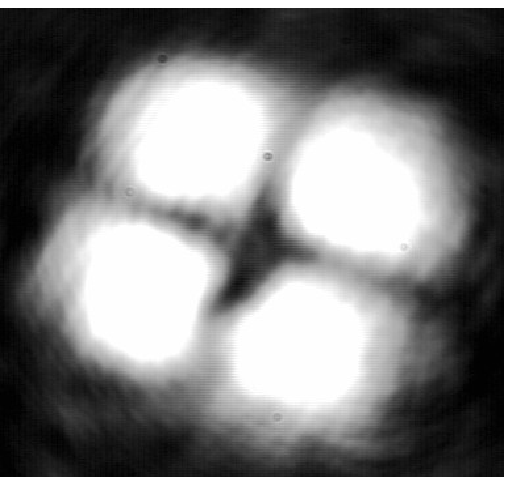}
\end{minipage}
\begin{minipage}[c]{0.2\linewidth}
\centering
\includegraphics[width=0.95\textwidth]{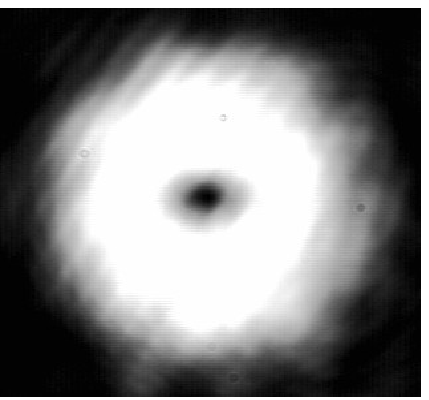}
\end{minipage}
\caption{Measured images of modes of $60\,\mathrm{\mu m}$
micro-cavity. The modes are HG00, HG01, HG11 and LG01, from left to
right, respectively.} \label{Fig. 7}
\end{figure}

We measured the transmission versus laser wavelength for a cavity
containing a layer of QDs. The transverse-mode frequency-spacings
become smaller as we approach the hemispherical limit by making the
cavity longer. Our results are consistent with predictions for the
hemispherical limit, paraxial-mode theory \cite{Siegman86}, which
predicts degenerate sets of modes, separated by $c/4\mathrm{L}$,
where L is the cavity length.

Figure \ref{Fig. 8} shows two scans over the range of wavelength
where the QDs absorb (750--760 nm). The finesse is about 50 at room
temperature. When we do a similar scan near 780 nm, where there is
less absorption, the finesse increases to 200. This is an indication
that we are observing cavity-enhanced absorption by the QD layer.
The predicted finesse is 600 based on reflectivity measurements of
the mirror alone. The lowered finesse is likely due to residual
contamination in the mirror dimple.

\begin{figure}[htbp]
\centering
\begin{minipage}[b]{0.45\textwidth}
\centering
\includegraphics[width=0.95\textwidth]{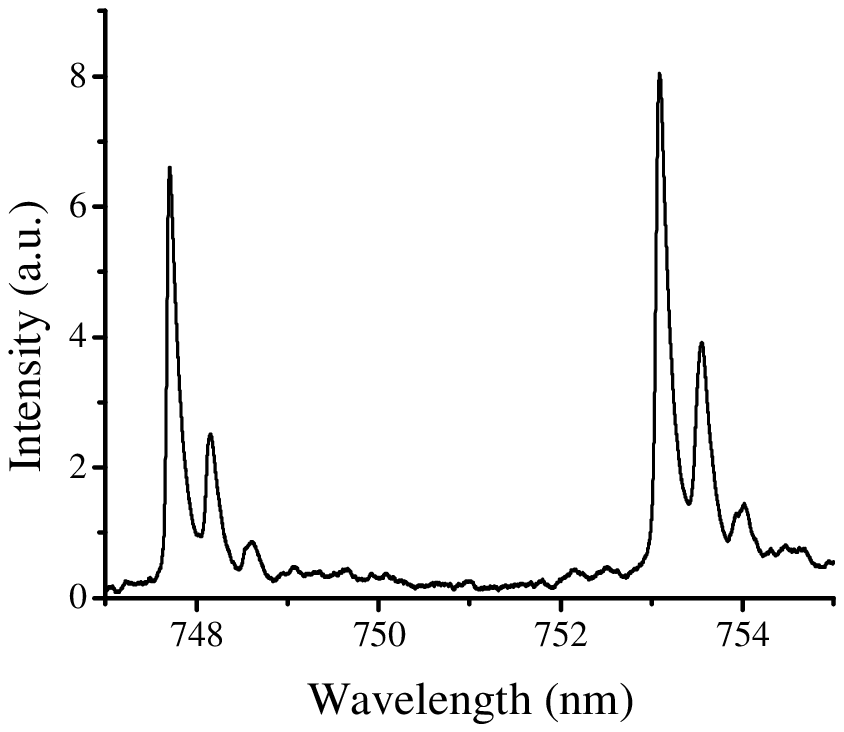} \\
 {Short Cavity}
\end{minipage}
\begin{minipage}[b]{0.45\textwidth}
\centering
\includegraphics[width=0.95\textwidth]{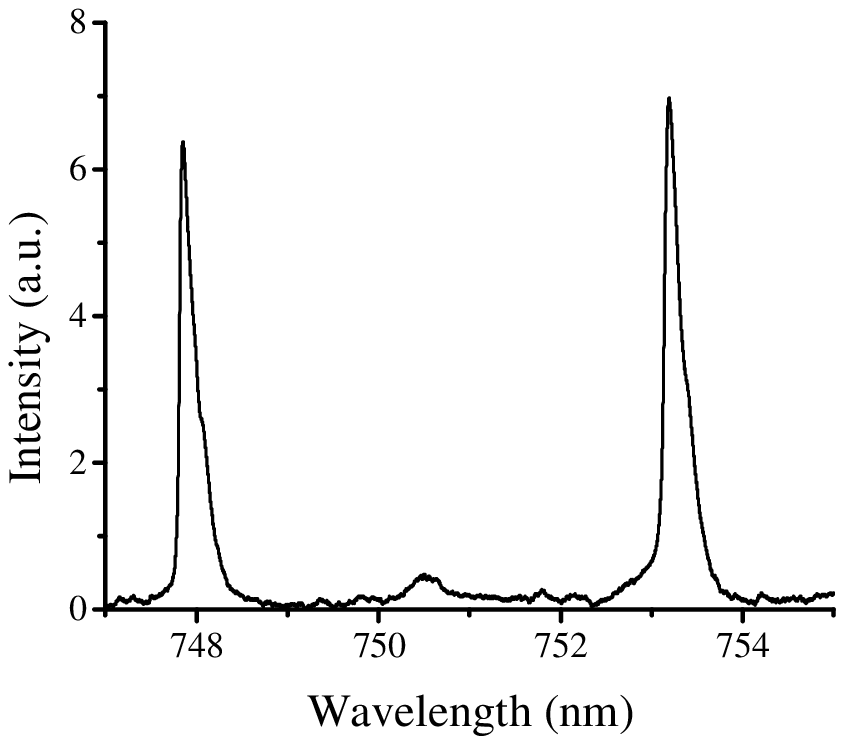} \\
 {Hemispherical Limit}
\end{minipage}
\caption{$60\,\mathrm{\mu m}$ cavity transmission spectra with QDs
at antinode. The cavity finesse is about 50 at room temperature.}
\label{Fig. 8}
\end{figure}

\subsection{Modeling the micro-cavity modes}

The QD-mode coupling strength is proportional to the amplitude of
the normalized cavity mode at the location of the QD. In order to
make the coupling very strong, it is necessary to localize highly
the transverse extent of the mode function in the vicinity of the
QD, and align the mode polarization vector with the dipole
transition matrix element of the QD. Determining the precise degree
to which this localization is possible is nontrivial, since the mode
structure for such a small cavity is non-paraxial, is non-separable
into polarization components, and is non-separable into longitudinal
and transverse modes \cite{Foster04}.

We have taken two approaches to modeling the modes of the
near-hemispherical micro-cavity. The two approaches are a fully
numerical one---finite-difference-time-domain (FDTD)
\cite{Pelton02}, and a hybrid analytic-numerical method
\cite{Bhongale03}. The computations account fully for the
distributed nature of the planar DBR mirror, an important aspect
since plane waves of different incident angles undergo different
phase shifts upon reflection there. The curved mirror is treated as
a perfect reflector, an approximation expected to be adequate since
the mode wave fronts are well matched to the mirror curvature. An
example of the FDTD method, showing the calculated energy density of
the mode versus position, is shown in Fig. \ref{Fig. 9}. The
calculations show that even in the presence of the DBR
angle-dependent phase shifts, the mode waist in the non-paraxial
regime is smaller than one wavelength.

An interesting result of the hybrid analytic-numerical method is a
novel DBR-induced spin-orbit coupling of modes, which leads to small
frequency splitting previously not identified \cite{Bhongale03}. The
method also predicts a spatial splitting of the fundamental Gaussian
mode (and other Gaussian modes) into a non-axis-symmetric inverted
``V'' shape.

\begin{figure}[htbp]
\centering
\includegraphics[width=0.4\textwidth]{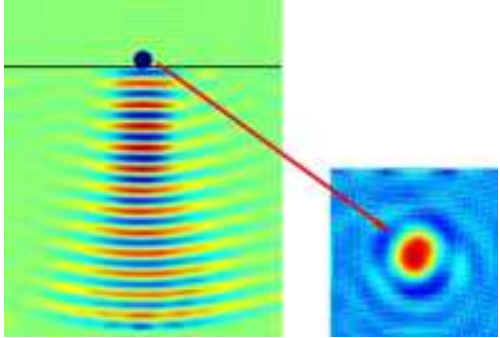}
\caption{Numerical model for micro-cavity mode energy density, where
the planar DBR structure is at the top and the curved mirror is in
the lower half of the figure. The QD sits in a bright local maximum
region in the first layer of the DBR. The results indicate that the
mode waist is of the order of one wavelength.} \label{Fig. 9}
\end{figure}

\section{Application}

The hemispherical micro-cavity we have fabricated has an excellent
prospect to achieve both strong coupling and efficient generation of
single- and pair-photons on demand. The hemispherical design is
geometrically stable with the only loss (other than surface scatter)
being by transmission through the end mirrors, not by diffraction
losses as occurs in other micro-structures \cite{Vuckovic02}. The
use of a concave micro-mirror with high-reflectivity over a
high-solid-angle makes the mode waist size at the planar DBR
diffraction limited and consequently leads to a large coupling
strength. It enables a direct out-coupling of the spontaneously
emitted single photons into a single-mode traveling wave, which is
highly desirable for the efficient and on-demand single-photon
generation. In addition, our system uses a cavity with adjustable
length and a transversely movable focal region, allowing good
spatial and spectral overlap of QD resonances with high-Q cavity
modes.

\subsection{Cavity QED Strong Coupling}

Cavity-QED strong coupling occurs when the electric-dipole
interaction frequency between an atom or QD and a single, unoccupied
mode exceeds the energy decay rates of the composite system. The
signature of strong coupling is a frequency splitting in the laser
transmission spectrum approximately equal to the twice of the
coupling constant, so called the normal-mode-splitting, which arises
from the coherent interaction of two degenerate systems---the single
QD and the single cavity mode. Such splitting can be viewed as a
lifting of degeneracy.

IFQDs can have dipole moments as large as 60--100 Debye
\cite{Andreani99}, yielding a vacuum Rabi splitting of
$49-81\,\mathrm{\mu eV}$, assuming a cavity waist of 1 micron and a
cavity length of 50 microns. As required for strong coupling, this
projected splitting would exceed the sum of the oscillator
dissipation linewidth, typically $15\,\mathrm{\mu eV}$, and the
cavity dissipation linewidth, $8\,\mathrm{\mu eV}$ for a length of
50 microns and a reflectivity of 0.996.

The transmission of an empty Fabry-Perot cavity has a series of
single peaks with high transmission at each resonance. In the
strong-coupling regime one of the peaks splits into two peaks, the
one in resonance with the QD transition, with a minimum located at
the position of former peak. This shows a strong enhancement of
system absorption at resonance. This interaction is suitable for
coherent quantum engineering concepts such as those being developed
in attempts to achieve quantum-information processing
\cite{Pellizzari95, Cirac97}.

\subsection{Photons on Demand}

Another important application of such strongly coupled cavity-QD
systems is the deterministic generation of single photons
\cite{Kuhn02, Kiraz03, McKeever04} or of photon pairs on demand
\cite{Stace03}. Such sources have wide applications in the emerging
field of quantum information science \cite{Bouwmeester00}. This is
particularly true for quantum cryptography, in which an essential
element of secure quantum key distribution (QKD) is an optical
source emitting a train of pulses that contain one and only one
photon \cite{Bennett92}. For example, a source having zero
probability for generating two or more photons in a pulse and
greater than 20\% probability of generating one photon would lead to
a great advance in QKD in daylight through the atmosphere
\cite{Hughes02, Resch05, Peng05}.

The high quality of the concave mirror substrate in our design opens
the possibility for very high cavity finesse. Currently we are
working to design a 99.95\% reflectivity coating, which should
achieve a finesse approaching 6,300. The unknown in this is how
smooth and regular the coating can be, when applied using standard
beam coating in such a small dimple. MBE-growth technique will
enable one to grow two QWs with a several-nanometer separation, with
a large enough barrier potential to prevent electron tunneling,
where IFQDs formed in each QW are each doped with an excess
electron. Quantum information processing can be implemented using
this structure for qubit storage and gate operation \cite{Pazy03},
while cavity modes used for transferring quantum information between
pairs of QDs. The cavity design should also lend itself to
applications in atomic quantum optics \cite{Mabuchi02} as well as
semiconductor optics \cite{Imamoglu99}.

\section*{Acknowledgments} This work was supported by the National
Science Foundation grant no. ECS-0323141 and by the Army Research
Office grant no. DAAD19-99-1-0344.

\end{document}